\documentclass[%
 reprint,
showpacs,
 amsmath,amssymb,
 aps,
]{revtex4-1}

\usepackage{graphicx}
\usepackage{dcolumn}
\usepackage{bm}

\begin{document}

\preprint{APS/123-QED}

\title{Rapid Information Transfer in Networks with Delayed Self Reinforcement}

\author{Santosh Devasia}
 \email{devasia@uw.edu}
\affiliation{%
 University of Washington \\
 Seattle, WA 98195-2600, USA
}%

\date{\today}

\begin{abstract}
The cohesiveness of  response to external stimuli depends on rapid distortion-free information transfer  across the network. 
Aligning  with the  information from the network has been used to model such information transfer. 
Nevertheless, the rate of  such diffusion-type, neighbor-based information transfer is limited by the  update rate at which each individual can sense and process information. 
Moreover,  models of the diffusion-type  
information transfer do not predict the superfluid-like  information transfer observed in nature. 
The contribution of this article is to show 
that self reinforcement, where each individual augments its neighbor-averaged information update using its previous update,  
can (i)~increase the information-transfer rate without requiring an increased, individual update-rate; and (ii)~capture the observed superfluid-like information transfer.
This improvement in the information-transfer rate without modification of the network structure or increase of the bandwidth of each agent 
can lead to better understanding  and design of networks with fast response.
\end{abstract}

\maketitle

\section*{Introduction}
The speed of information transfer across the network can impact the cohesiveness and effectiveness of the network's response to external stimuli. 
Aligning with the information from a network has been used to model a range of information transfer in nature  such as 
information diffusion in social networks~\cite{Nematzadeh_PRL_14,Ruan_PRL_15}, complex networks~\cite{Nicosia_PRL_2017,Czaplicka_PRE_16,Masuda_PRL_13}, 
and flocking dynamics, e.g.,~\cite{Nagai_15,Huth_92,Vicsek_95,Couzin_02,Makris_09,Bialek_12,Attanasi_14}. 
In general, a faster 
information-transfer rate  can be achieved by increasing the alignment strength, i.e., by scaling up the individual update that is based on information from neighbors. 
Nevertheless, such an increase in the alignment strength  will require 
an increase in the information-update rate, which is limited  by each individual's ability to sense and process external stimuli.
Hence there is a limit to the maximum rate of information transfer possible  with a fixed update rate.

The main contribution of this paper is to develop a self-reinforcement approach that can increase the information transfer without 
the need to change the network structure or the bandwidth (information-update rate) of the individual agents. Rather, the 
proposed approach uses delayed versions of the previous updates from the network to self reinforce the current update and improve the 
overall network response. 
Such faster response rate with  limited individual performance (update rate)~\cite{Berdahl_13} can influence current studies in group decision making, e.g.,~\cite{Strandburg_15,van_de_Waal_13}, models of cohesiveness in groups, e.g.,~\cite{Couzin_02,Attanasi_14}, and interactions between layers of networks~\cite{Nicosia_PRL_2017}, improve communication of engineered swarms such as robots~\cite{Halloy_07,Rubenstein_14}, and potentially lead 
to better understanding of  response to external stimuli in biological systems~\cite{Ioannou_12}. 
%

Another challenge with current models of the neighbor-averaged  
diffusive information transfer is that they do not predict the superfluid-like information transfer observed in biological flocking~\cite{Attanasi_14,Cavagna_15}.
Superfluid-like information transfer leads to undamped propagation of the radial acceleration across the flock, which is 
important to achieve equal-radius (parallel) trajectories for cohesive maneuvers~\cite{Attanasi_14}. 
Nevertheless,  superfluid-like models also require an increase in update rate for fast response. 
This article shows that current  diffusive models can be modified to capture the superfluid-like 
information transfer observed in nature without the need to increase the bandwidth of the individual agents. 
Since delays are available 
in neural circuits, the  delayed self-reinforcement (DSR) method might be potential mechanism to explain the superfluid-like observations.

\section*{Models  with and without DSR}  
The alignment of each individual $i$ based on the information available to its neighbors $N_i$ is modeled below. 
Let the new information $I_i(k+1)$ for the $i^{th}$ individual   be found from the  information update 
given by 
\begin{equation}
\left[ I_i (k+1) - I_i (k)  \right] = - K_s {\Delta}_i(k) \delta_t  + \beta \left [I_i (k)  - I_i (k-1) \right], 
\label{Eq_discrete_DSR}
\end{equation}
where different integers $k$ represent  the  update time instants $t_k = k\delta_t$, the time interval between updates  $\delta_t$ depends on the 
reaction time of the individual, $\beta$ is the DSR gain on the previous update, 
$K_s$ is 
the alignment strength, and ${\Delta}_i(k)$ is the average difference in the information between the individual and its  $| N_i |$ neighbors in the network 
\begin{equation}
{\Delta}_i(k) =   \frac{1}{| N_i |} \sum_{j \in N_i}  \left[ I_i (k)  - I_j(k)   \right] .  
\label{Eq_diffusion}
\end{equation}
In flocking, the network connections depend on metric distance or topological distance~\cite{Ballerini_08}.
In this model,  the set of neighbors $N_i$  also includes the information source $ I_s$ when the individual $i$ is a leader with direct access to the source. 
This model corresponds to the standard diffusion-based information update if the DSR gain $\beta$ in Eq.~\ref{Eq_discrete_DSR} is set to zero, e.g., ~\cite{Huth_92,Vicsek_95}.

\begin{figure*}
\begin{center}
    \begin{tabular}{@{}cccc@{}}
        \includegraphics[width=0.18\textwidth]{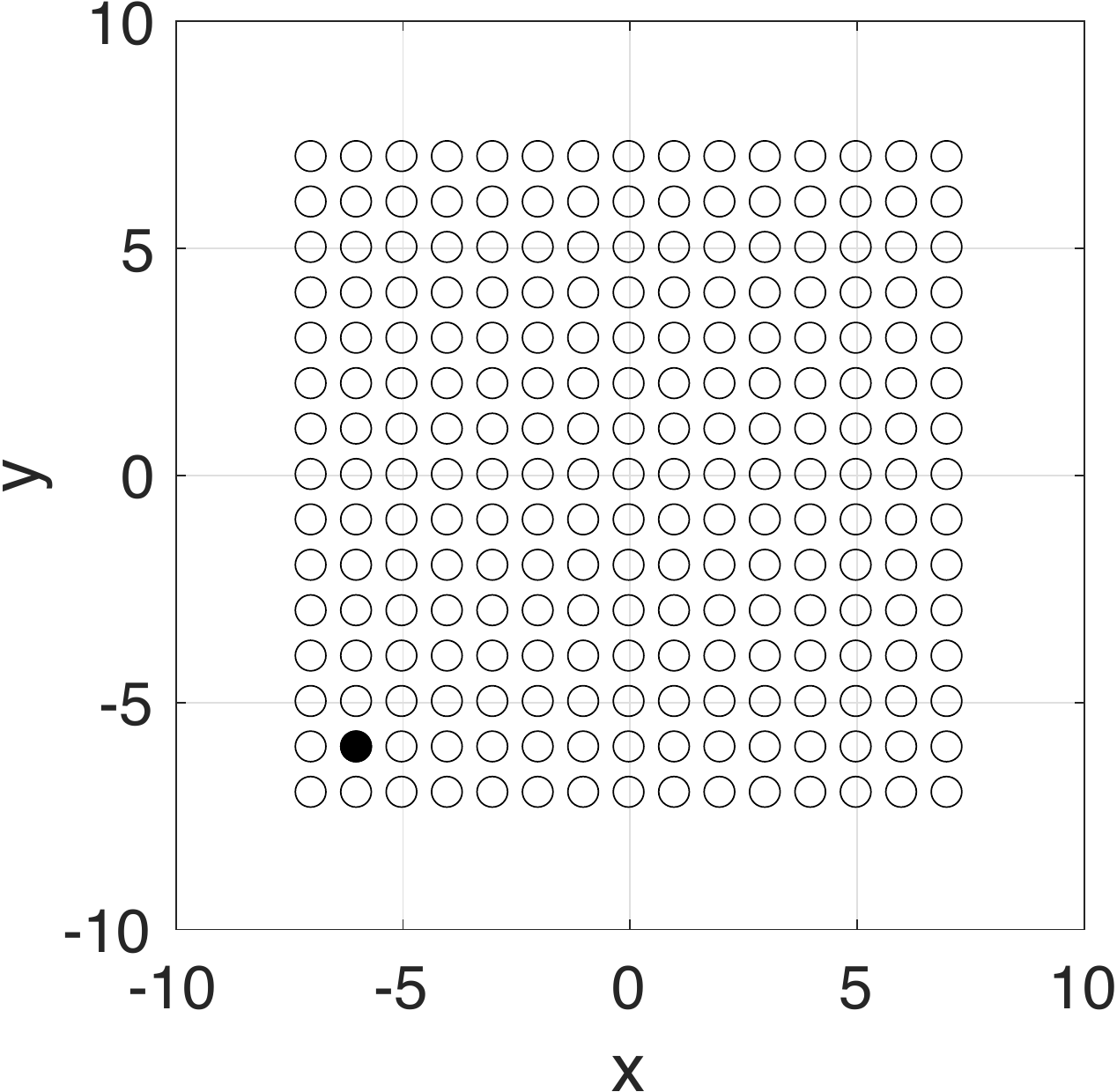}  & 
    \includegraphics[width=0.25\textwidth]{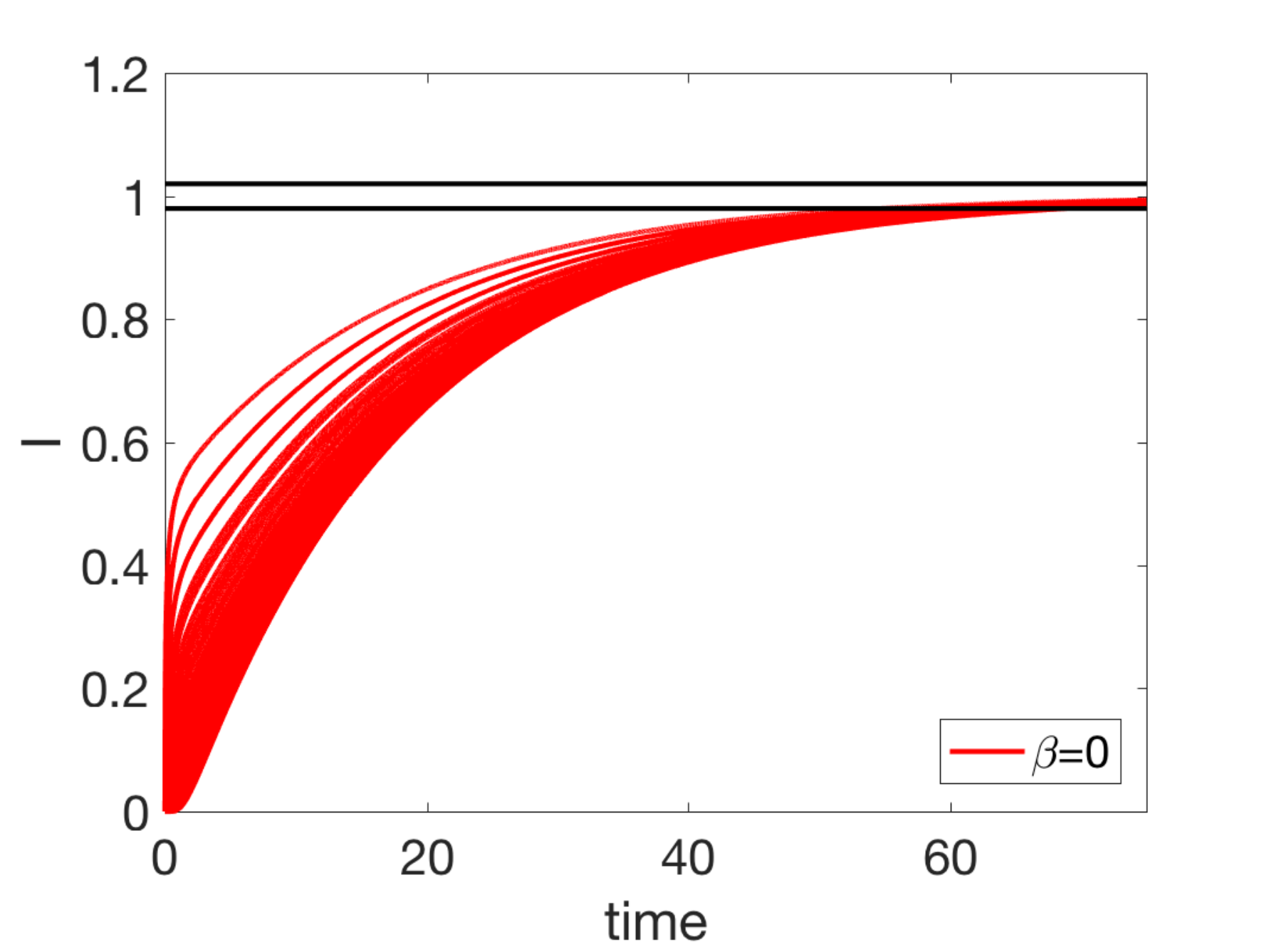}  & 
     \includegraphics[width=0.23\textwidth]{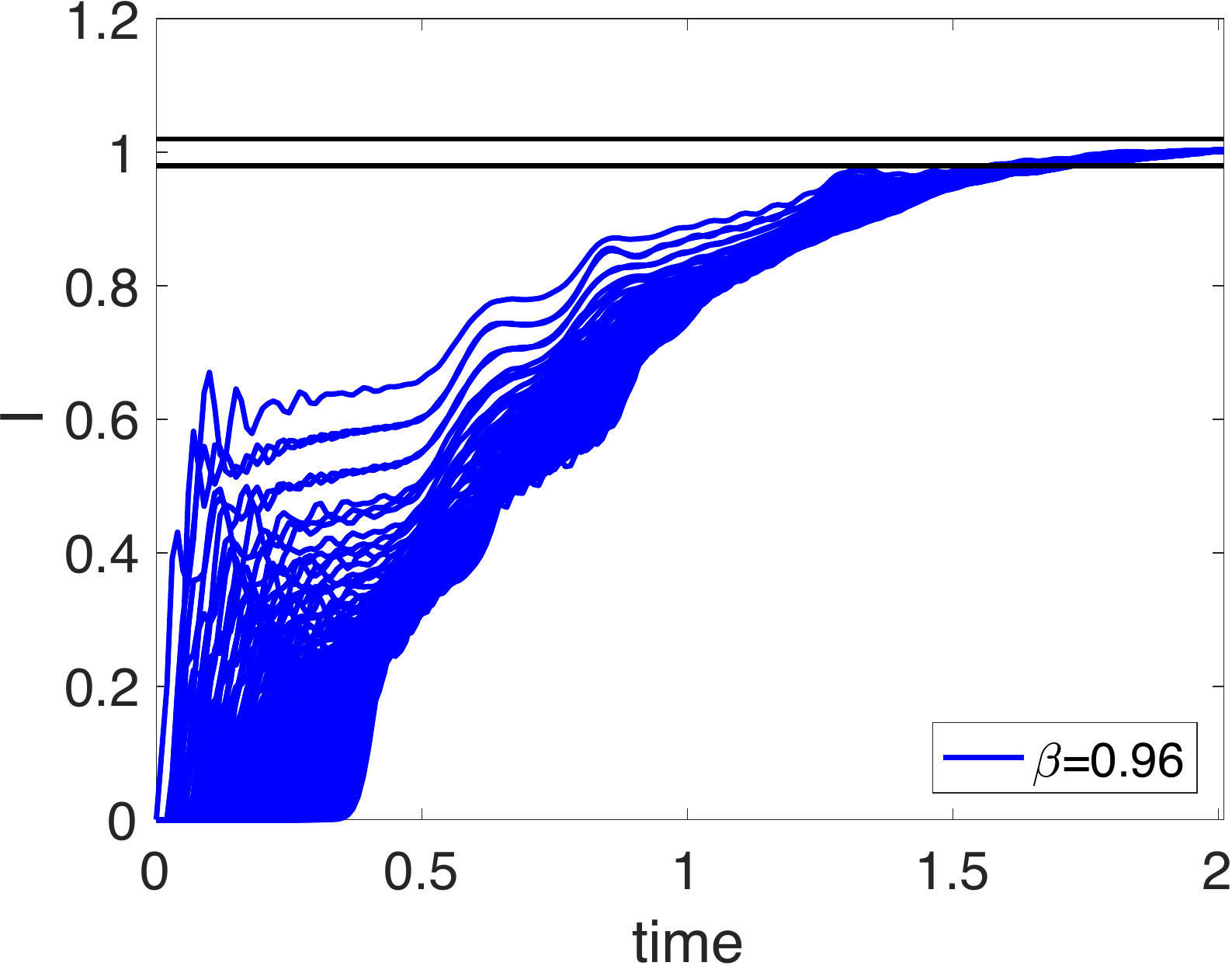}  & 
          \includegraphics[width=0.23\textwidth]{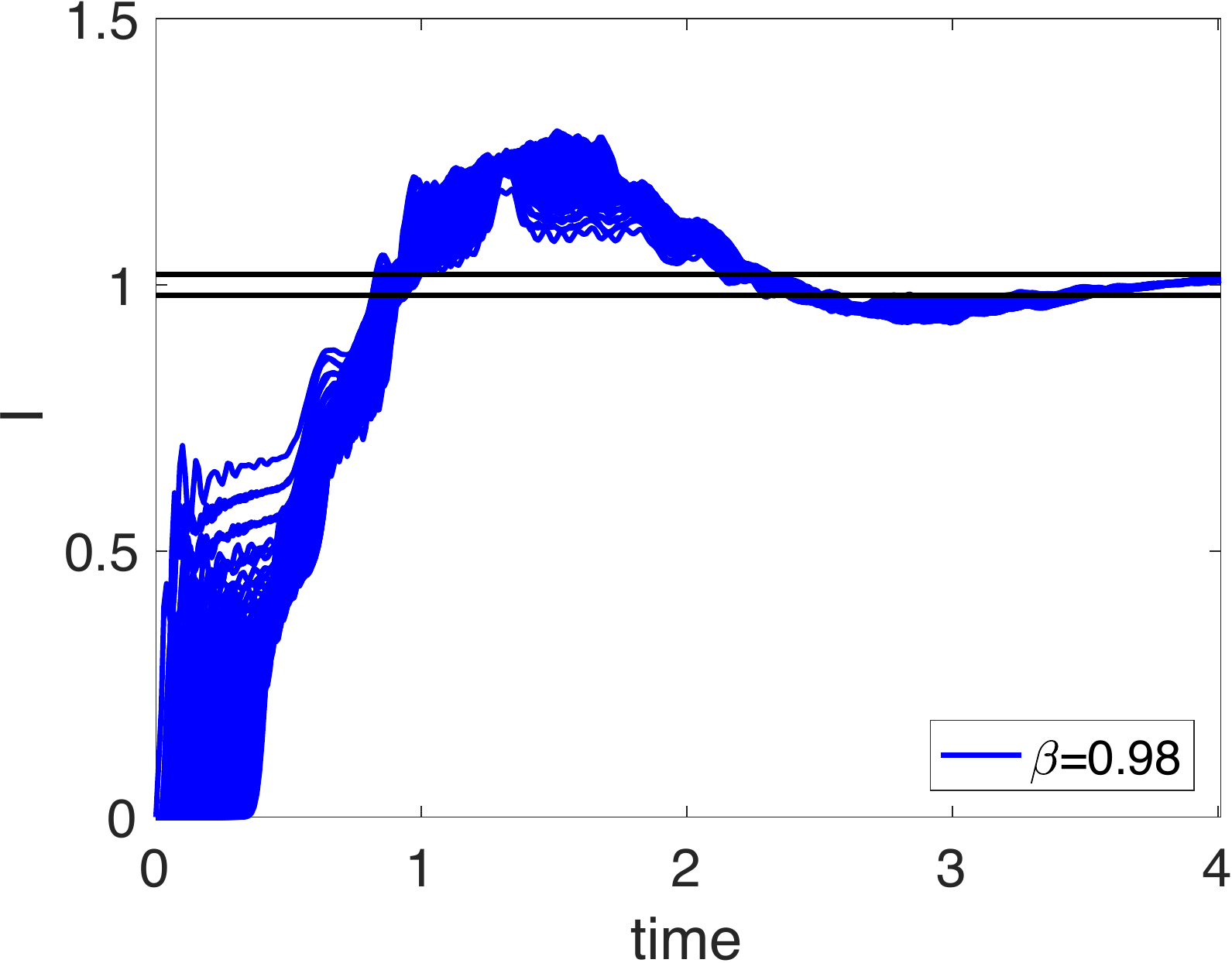} \\
          ~~~~~~(a) & ~~~~~~(b) & ~~~~~~(c) & ~~~~~~(d) 
  \end{tabular}
  \end{center}
  \vspace{-0.2in}
\caption{
Comparison of settling time $T_s$  needed for information $I$ to reach and stay within $\pm$ $2\%$ of the final value of one. 
(a) The initial configuration of the agents. 
(b): without DSR, i.e.,  $\beta=0$   leads to settling time $T_s=69$s;  (c): use of DSR with  gain $\beta=0.96$ leads to smaller settling time 
$T_s=1.72$s; and (d): a larger DSR gain $\beta=0.98$ leads to oscillations and a larger settling time $T_s=3.52$s. 
%
}
\label{fig_1_main}
\end{figure*}

\vspace{0.1in}
\paragraph*{Information transfer improvement with DSR}  
For a given system  update time $\delta_t$, 
DSR can lead to substantial performance improvement when compared to the standard diffusive information transfer without the DSR as illustrated in Fig.~\ref{fig_1_main}. 
%
The  system comprised of 
$225$ individuals placed in a $25\times 25$ regular array,  where the spacing in the $x$ and $y$ 
direction was $1$ m. The neighborhood $N_i$  of each individual 
was considered to be a disc of radius  $r=1.2$m from the individual $i$. 
Thus, the average distance of individuals in the neighborhood was $a=1$m. 
The leader is the individual shown as a solid black dot in Fig.~\ref{fig_1_main}(a). 
The initial value of the source information and  of all the individuals were zero. The source information  is switched to one at the 
start of the simulations. Without DSR, the information 
transfer becomes unstable as the alignment strength is increased to $K_s=101$ from $K_s=100$. 
Therefore, the alignment strength  was selected to be high ($K_s=100$) to enable fast response, but smaller than the value that causes instability. 
Then, the DSR gain was selected to yield a fast response; it was varied from $0$ to $0.99$ in increments of $0.01$ 
and the gain of $\beta=0.96$ was selected as it yields a fast response without substantial overshoot.
%
%
%
With the DSR gain selected to avoid oscillations, $\beta=0.96$, the DSR leads to a 
substantial  (more than an order) 
reduction in the settling time (i.e., the time needed for all the individual responses 
to become close (within 2\%)  and stay close to the maximum value of the source information) ---
from $69$s to $1.72$s.
Without DSR, similar substantial improvements are not possible 
since increases in the alignment strength 
leads to instability.  
Thus, for a given update rate, 
the proposed DSR leads to better transfer of rapidly changing  information when compared to the standard case without the DSR.

\section*{Impact  of DSR on flocking}  
DSR can improve the cohesiveness of flocking maneuvers,   when  the 
orientation of each individual is considered to the be the information $I$ being transferred using local alignment to neighbors. 
To illustrate, the  position components $x_i,$, $y_i$ 
of each individual is 
updated as 
\begin{equation}
\begin{array}{rcl}
x_i (k+1) & = &    x_i (k)  + v \delta_t \cos{I_i},  \\
 y_i (k+1) & = &   y_i (k)  + v \delta_t \sin{I_i}, 
 \end{array}
\label{Eq_flocking}
\end{equation}
where $v$ is the fixed speed of each individual. To focus on the impact of orientation-information transfer on the maneuver, other effects such as
speed changes  or strategy changes to maintain spacing  between individuals or density are not included in the simulations, e.g., as studied in~\cite{Gallos_PRL_08,Nematzadeh_PRL_14,Nicosia_PRL_2017,Couzin_02,Gregoire_04,Buhl_06,Herbert_11,Cavagna_13}. 
Note that the set of neighbors can change during these 
simulations, however, the initial spacing is selected to ensure that each individual has at-least two neighbors at the start of the simulations.

\begin{figure*}
\begin{center}
    \begin{tabular}{@{}cccc@{}}
    \includegraphics[width=0.23\textwidth]{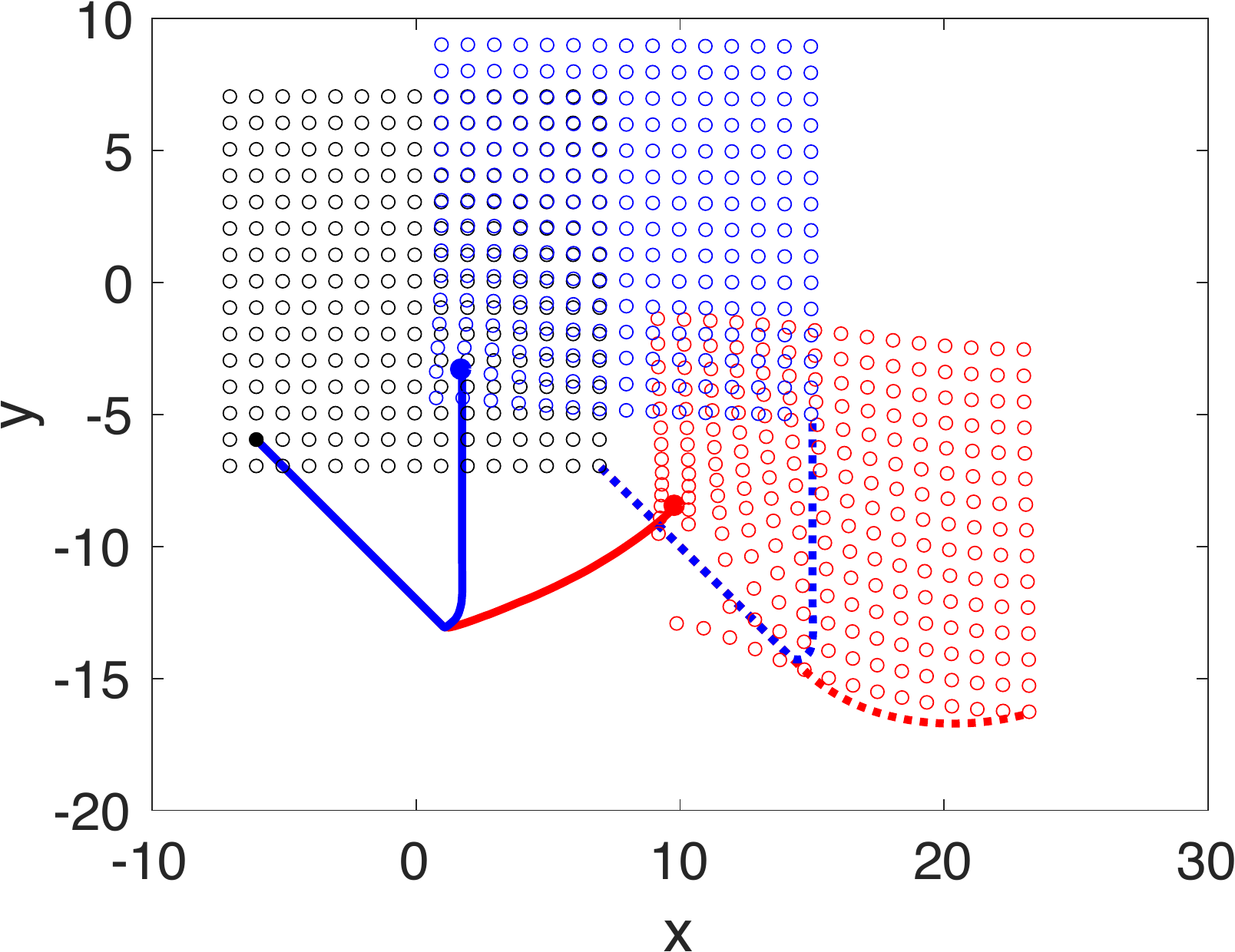}  & 
     \includegraphics[width=0.23\textwidth]{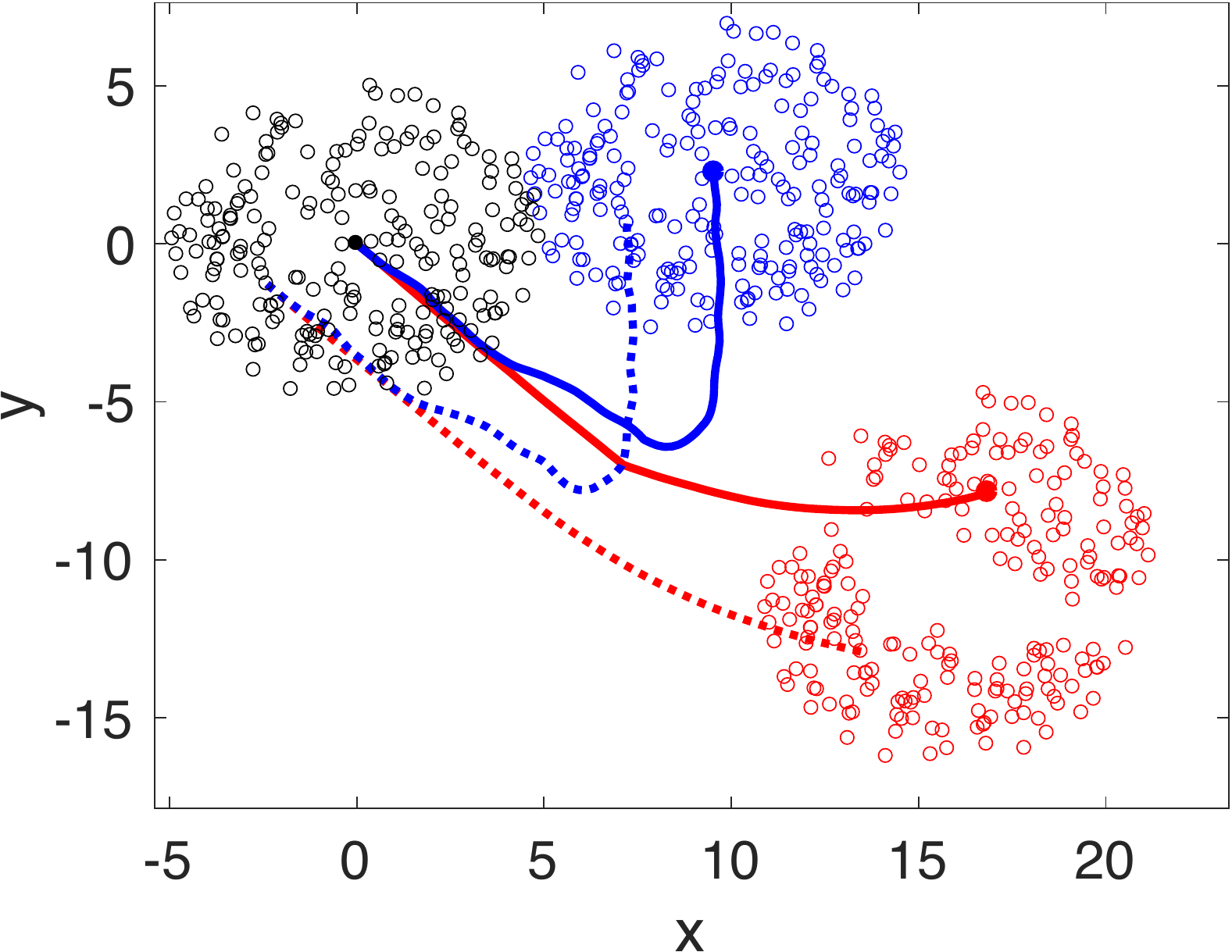}  & 
     \includegraphics[width=0.23\textwidth]{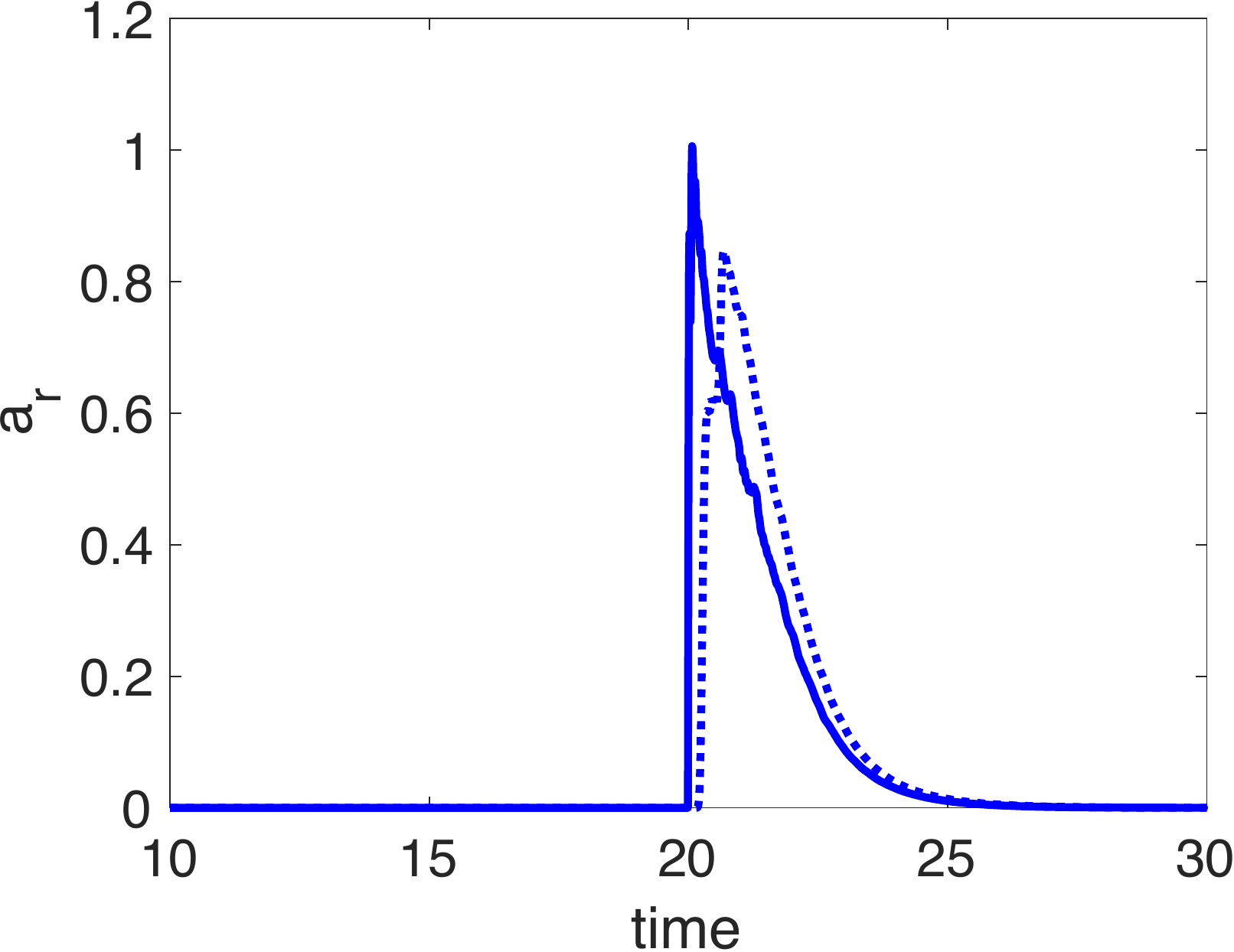} & 
      \includegraphics[width=0.24\textwidth]{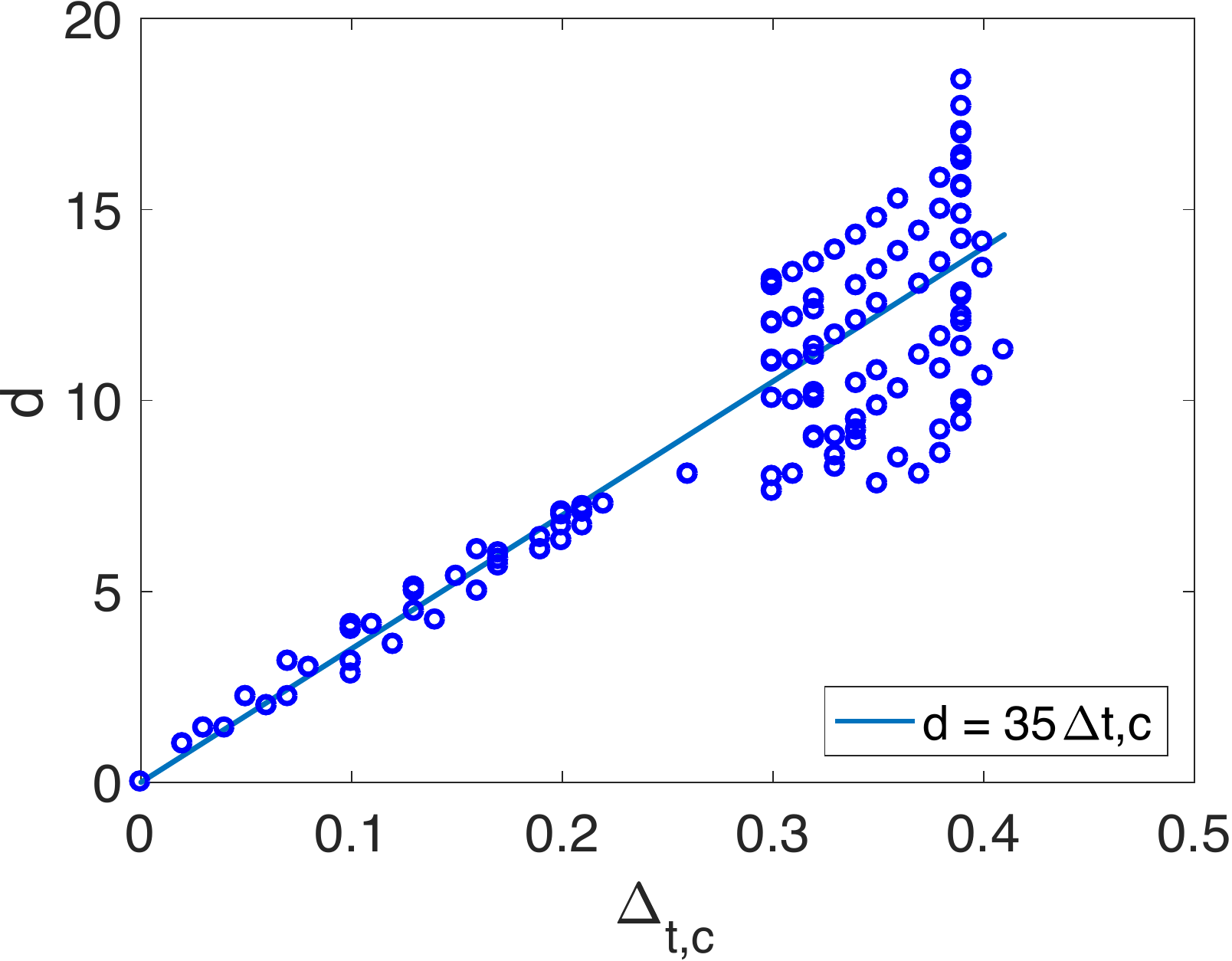}  \\
          ~~~~~~(a) & ~~~~~~(b) & ~~~~~~(c) & ~~~~~~(d) 
  \end{tabular}
  \end{center}
  \vspace{-0.2in}
\caption{
Comparison of cohesiveness of a  constant-speed $v$ turn maneuver.
Videos (V1-V4)  are available in supplementary materials.
Cohesiveness is indicated by comparing 
the movement of the leader (solid line) and an individual farther away (dotted line); and 
the initial (black dots) and final formations (blue with DSR, red without DSR). The solid dot represents the leader which has direct access to the information source $I_s$.
(a) Uniformly initial spacing without noise in the  updates. 
(b) Random initial spacing (black dots) in a circle and  with noise in updates. 
(c) The radial acceleration $a_r$ of the  leader (solid line) and individual farther away (dotted line).  
(d)  
The impact of  distance $d$ from the leader on the 
shift in time $\Delta_{t,c}$ needed to correlate the radial acceleration of each individual to that of the leader.
}
\label{fig_2_main}
\end{figure*}

The use of DSR leads to improved cohesiveness in  maneuvers when compared to the case without DSR, as illustrated by the results in Fig.~\ref{fig_2_main}. 
The desired information source $I_s$ (i.e., the desired orientation of the agents) is switched from an initial value of  $-\pi/4$ to the final value of $\pi/2$. 
Two cases, uniform and random initial distribution, of the agents are considered, and in each case, the simulations were performed with and without the noise in the update. 
For the random case as in Fig.~\ref{fig_2_main}(b), the initial locations were randomly chosen in a disc of radius $r_d = 25/3$, which was selected to  be small enough to 
ensure that there was at-least 
two individuals in each neighborhood $N_i$. The radial distance  $r_i$ from the center was chosen to be the square root of a uniformly-distributed 
random variable between $0$ and $r_d$  and the angle $\theta_i$ 
was selected to be randomly distributed 
between $0$ and $2\pi$ radians to yield the initial locations as $x _i = r_i \cos(\theta_i) $ and $y_i = r_i\sin(\theta_i) $.
Moreover, a uniformly-distributed random noise (between $-0.025$ rads and $0.025 $ rads) was added to the estimates of the  averaged-neighbor 
orientation information update in Eq.~\ref{Eq_discrete_DSR}.

The maneuver with DSR is more cohesive,  for both  uniform and random initial distribution, as seen in the 
similarity of the initial and final formations when compared to the case without the DSR, and also seen in the Videos V1-V4.  
Even with the addition of noise in the information update, the overall motion remains cohesive, see Video V3. 
In Fig.~\ref{fig_2_main}, the turn movement (blue solid line)  of the leader  is similar to that of an 
individual which is farther away, which is an important feature in biological flocks which exhibit equal-radius (parallel) trajectories~\cite{Attanasi_14}. 
In contrast, without DSR, the final direction of the leader (slope of the solid  red line) is different from that of individuals farther away. Moreover, the slower transfer of turn-information leads to a larger turn radius without the DSR when compared to the case with the DSR. 
The time shift $\Delta_{t,c}$  needed for the individual radial acceleration to best correlate with the radial acceleration of the leader varies linearly with distance $d$ from the leader (for individuals 
close to the leader), with the DSR approach, as seen in Fig.~\ref{fig_2_main}(d). 
The overall speed of information transfer across the network is  $47$ m/s, where the correlation time 
delay $\Delta_{t,c}$ is $0.389$ s for a distance of $18.38 $ m. 
Moreover,  the magnitude of the radial acceleration does not reduce 
substantially with distance from the leader, as seen in Fig.~\ref{fig_2_main}(c). 
Both these features, linearity of the information transfer with time and low distortion,  
are indicative of superfluid-like flow of information observed in nature that 
cannot be explained by standard diffusion models~\cite{Attanasi_14}. 
Thus, the proposed DSR 
captures the superfluid-like turning maneuvers observed in nature.

\section*{Superfluid-like behavior with DSR} 
To understand the impact of the DSR gain $\beta$ selection on capturing the superfluid-like behavior in the results in Fig.~\ref{fig_1_main}, 
the   information update in Eq.~\ref{Eq_discrete_DSR} 
is first rewritten as 
\begin{equation}
\begin{array}{rcl}
\frac{\beta}{\delta_t} \left\{ \left[ I_i (k+1) - I_i (k)  \right] - \left[ I_i (k) - I_i (k-1)  \right]   \right\}
& & \\ + 
\quad \frac{1-\beta}{\delta_t}   \left[ I_i (k+1) - I_i (k)  \right]  & & \\
 = &  - K_s {\Delta}_i(k),  & 
\end{array}
\label{Eq_DSR_Rearrangement}
\end{equation}
and then  approximated, 
when the update interval $\delta_t$ is small compared to the information-transfer response, as
\begin{equation}
\beta \delta_t \frac{d^2}{dt^2} I (t)   +(1-\beta) \frac{d}{dt} I (t) = \frac{a^2}{4} K_s  \nabla^2 I (t)      
\label{Eq_approximation_DSR}
\end{equation}
where  $a$ is the average distance to the neighbors and $ \nabla^2$ represents the Laplacian. 
This approximate model captures a  broad set of behaviors. 
As the DSR gain  
tends to one,  the damping  term $(1-\beta)$ tends to 
zero and the overall behavior changes from overdamped (e.g., $\beta=0$) to being critically damped (e.g., $\beta=0.96$) to oscillatory undamped (e.g., $\beta=0.98$), 
as seen in Fig.~\ref{fig_1_main}. Large oscillations  can lead to distortions in the information loss, and ideally the DSR gain $\beta$ is tuned to be close to critical damping. 
For small DSR gain 
the DSR dynamics approximates the overdamped standard diffusion-type information transfer 
\begin{equation}
 \frac{d}{dt} I (t) =  \frac{a^2}{4}  K_s  \nabla^2 I (t).     
\label{Eq_DSR_diffusion_limit}
\end{equation}
With a larger DSR gain,  
the DSR dynamics approximates the superfluid-type information transfer, i.e., 
\begin{equation}
 \frac{d^2}{dt^2} I (t)   = \frac{ a^2 K_s }{ 4 \delta_t}  \nabla^2 I (t) ~ = c^2  \nabla^2 I (t) 
\label{Eq_DSR_superfluid_limit}
\end{equation}
where a smaller update time $\delta_t$ (which is possible if the individuals can respond faster) leads to a larger speed of information propagation $c$.

Both the standard diffusive model and the second-order superfluid-type model in Eq.~\ref{Eq_approximation_DSR}  can achieve faster information transfer, similar to the 
case with the use of DSR, as seen in 
Fig.~3.  
The superfluid-like simulations, were computed based on Eq.~\ref{Eq_approximation_DSR}
as 
\begin{equation}
\begin{array}{rcl}
I (k+1)  & = & I(k) + \dot{I}(k) \hat{\delta}_t \\
\dot{I}(k+1) & = & \dot{I}(k) - \frac{(1-\beta) }{\beta \delta_t}\dot{I}(k) \hat{\delta}_t  
+ \frac{K_s }{\beta \delta_t}   {\Delta}_i(k) \hat{\delta}_t
\end{array}
\label{Eq_approximation_DSR_simulation}
\end{equation}
where the update rate was $\hat{\delta}_t = 1.246 \times 10^{-4}$ s.
The settling time with the standard diffusive model is $1.72$ s  (with the alignment strength $K_s$ increased about $40$ times, from $100$ to $4011$) and with the superfluid-like model in Eq.~\ref{Eq_approximation_DSR}  is $1.78$ s, which are similar to the settling time 
of $1.72 $ s with the DSR.  However, the standard diffusive model requires a proportional decrease in update time by about $40$ times, from $0.01$ s to $ 2.49 \times 10^{-4}$ s to maintain stability. 
With the same update time of  $ 2.493 \times 10^{-4}$ s,  the superfluid-like model was unstable, and hence, the results  in Fig.~3 are shown with the update time reduced by half, i.e., to 
$ 1.246 \times 10^{-4}$ s. Note that the information transfer distance is linear in time $t$ with the DSR. 
In contrast, the  information transfer distance $d$ is proportional to the square root of time $\Delta_t$ with the standard diffusion model.
The speed of information transfer with DSR is close to the expected value for the superfluid case from the 
expression of $c$   in Eq.~\ref{Eq_DSR_superfluid_limit}.  In particular, with an average distance of $a=1$, 
the predicted speed $c$ in Eq.~\ref{Eq_DSR_superfluid_limit} is $c=50$ m/s.  This is close to the speed of information transfer seen in the results in Fig.~3, where 
information is transferred over a distance of $18.38$ m in  $ 0.39$ s, i.e., at a speed of $47$ m/s.  

\vspace{0.1in}
In summary, the use of the DSR neighbor-based alignment achieves the superfluid-type  information transfer, and increases the overall information transfer rate in the network without  requiring a corresponding increase in individual information-update rate.  In contrast, current superfluid-like model and standard diffusion models can only achieve the faster information transfer by increasing the 
individual, information-update rate.

\begin{figure*}
\begin{center}
    \begin{tabular}{@{}cccc@{}}
    \includegraphics[width=0.24\textwidth]{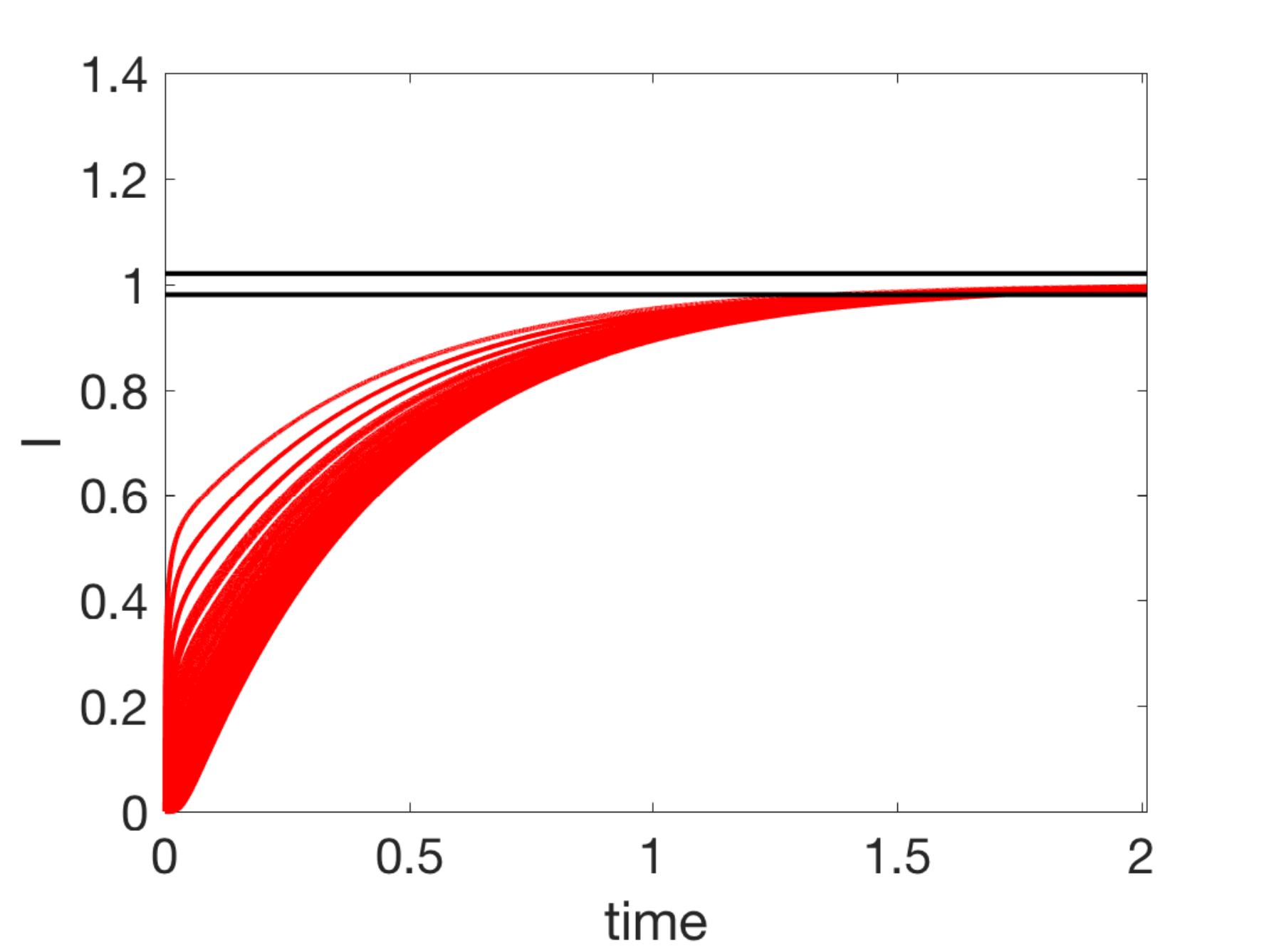}  & 
     \includegraphics[width=0.24\textwidth]{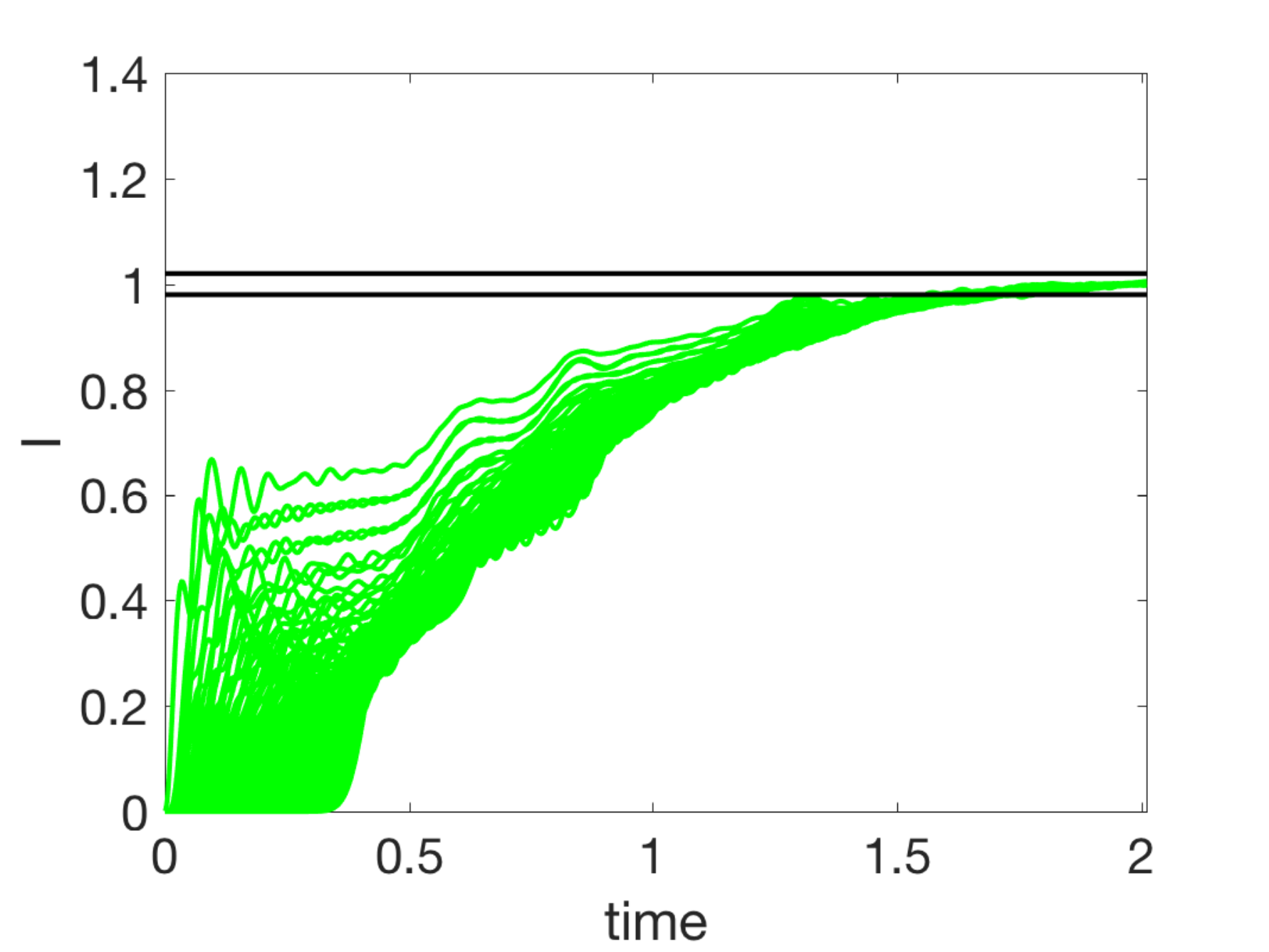}  & 
     \includegraphics[width=0.22\textwidth]{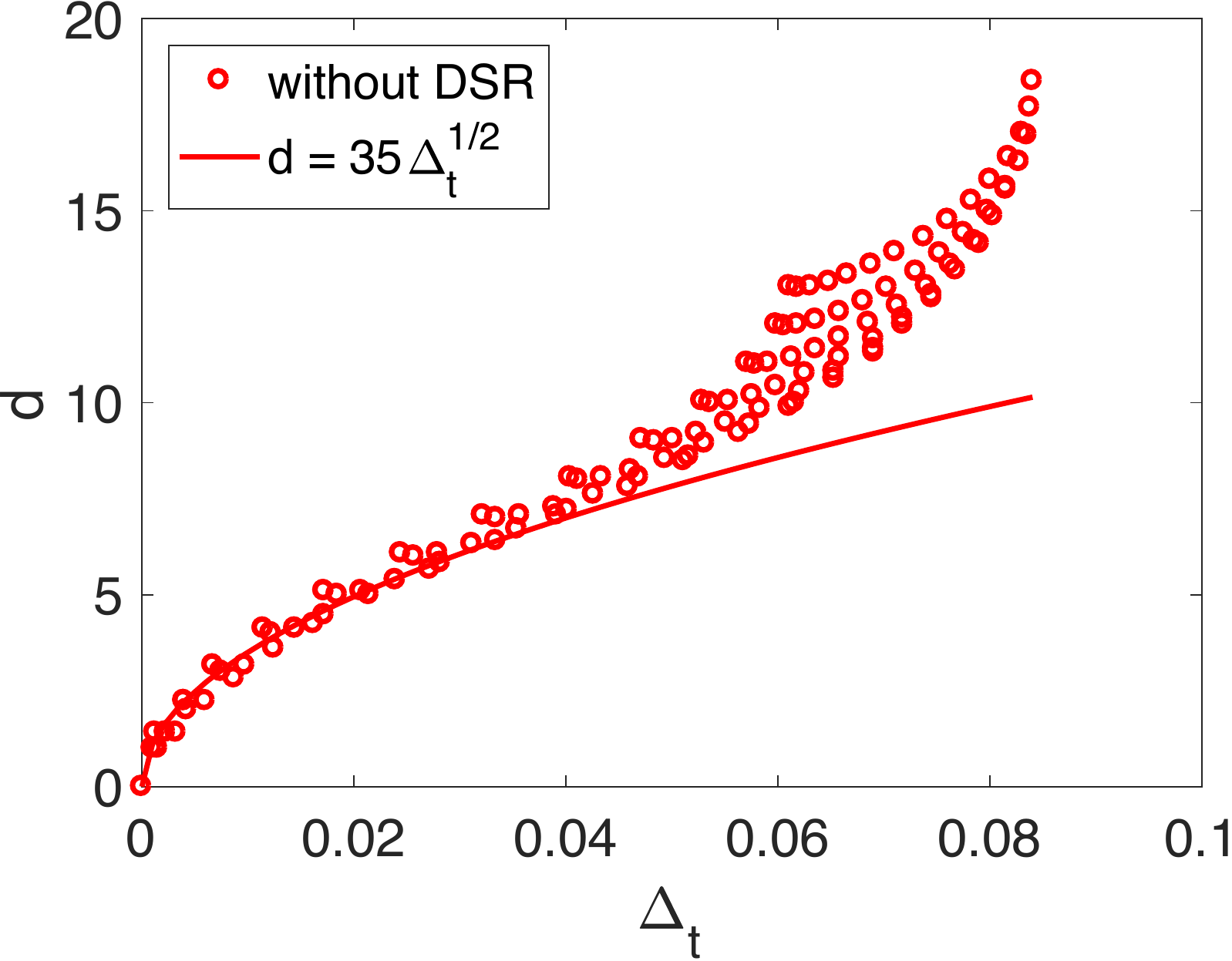} & 
      \includegraphics[width=0.22\textwidth]{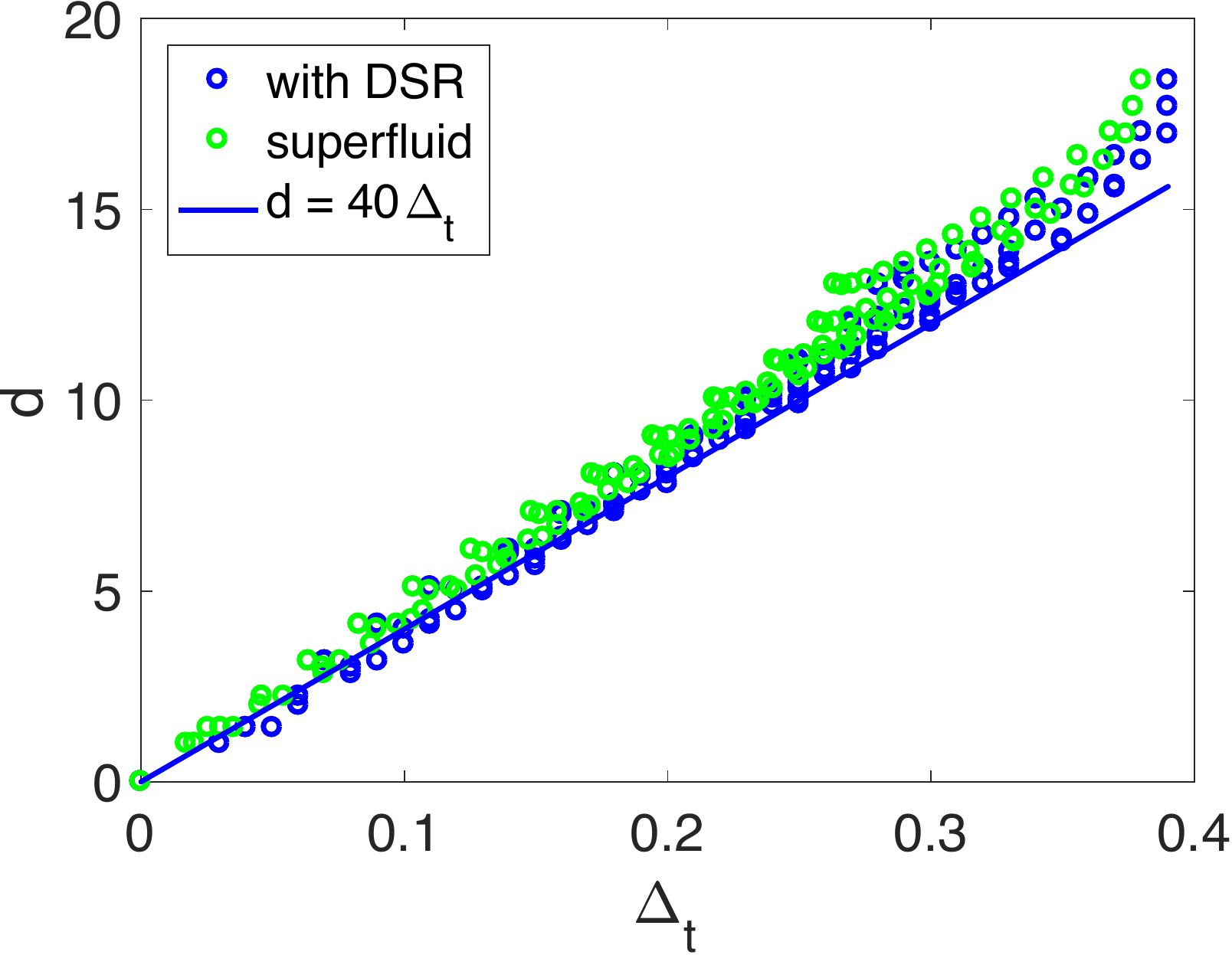}  \\
          ~~~~~~(a) & ~~~~~~(b) & ~~~~~~(c) & ~~~~~~(d) 
  \end{tabular}
  \end{center}
  \vspace{-0.2in}
\caption{
Information transfer similar to the case with DSR (Fig.~1(b)) is achieved 
with: (a) standard diffusion model 
by increasing the  alignment strength $K_s$  from $100$ to $4011$ and  
decreasing  the update time from $0.01$ s to $ 2.49 \times 10^{-4}$ s; and (b) with the 
superfluid-type model in Eq.~\ref{Eq_approximation_DSR}. 
The time delay $\Delta_t$ between the leader and other individuals to reach $0.1$ as a function of the distance $d$  from the leader: 
(c)~the information transfer distance $d$ is proportional to the square-root of time $\Delta_t$ for the
standard diffusive model without DSR  for individuals close to the leader and (d) linear for the DSR case and for the superfluid-like model.
}
\label{fig_3_main}
\end{figure*}

%

\end{document}